\documentclass{elsart41}
\usepackage{graphics}
\usepackage{graphicx}
\usepackage{amssymb}
\begin{document}

\begin{frontmatter}



\title{
Global Magnetic Phase Diagram and Local Quantum Criticality in Heavy
Fermion Metals}

%

\author[AA]{Qimiao Si},
\ead{qmsi@rice.edu}

\address[AA]{Department of Physics \& Astronomy, Rice University,
Houston, TX 77005--1892, USA}


\begin{abstract}
We address the global magnetic phase diagram of Kondo lattice systems.
Through 
the distinct Fermi surface properties of the various 
phases at zero temperature, 
we
argue that the phase diagram
supports two classes of quantum critical point. One of these 
describes 
a direct transition from a magnetic metal phase with
localized $f-$electrons to a paramagnetic
one
with itinerant
$f-$electrons. This
result
provides the context for the picture of local 
quantum criticality, in which the Fermi surface jumps across 
the quantum critical point and the quasiparticle residue 
vanishes as the quantum critical point is approached from
either side. Some of the unusual experiments, concerning 
the phases and quantum critical points of heavy fermion metals,
are discussed from
the present perspective. These developments have implications
in broader contexts. In particular, they form a part of the 
growing evidence for quantum criticality that goes beyond 
the orthodox
description in terms of order-parameter fluctuations.
\end{abstract}

\end{frontmatter}


The past decade has witnessed the modern era in the study of 
heavy fermion metals. In part due to cross fertilization with
studies on other classes
of strongly correlated
electron systems
such as high temperature
superconductors, the heavy fermions have
become
a prototype family of materials for non-Fermi liquid
behavior~\cite{Maple,Stewart} and quantum
criticality~\cite{Lohneysen,Mathur,Custers}. One advantage
that
the heavy fermions have is that antiferromagnetic quantum critical
points (QCPs) have been explicitly observed. The ensuing studies
of the 
quantum critical heavy fermions have helped 
establish the notion that quantum criticality
leads to
both unconventional superconductivity and non-Fermi liquid
behavior. Moreover, they have been instrumental in the growing 
realization that quantum criticality can be considerably more complex
than its classical counterpart.

The universal properties of most classical critical points are
described
in terms of spatial fluctuations of an order parameter, $m({\bf x})$.
It is conventional wisdom that a similar description applies to QCPs:
the only change,
reflecting the mixing of statics and dynamics,
is that the order parameter fluctuations, $m({\bf x}, \tau)$,
are now in both space and imaginary time~\cite{Hertz76,Pfeuty70,Young75}.
The experiments in heavy fermions have strongly contradicted this
picture~\cite{Stewart},
pointing to emergent critical modes that are beyond the
order-parameter 
fluctuations.

We have addressed this issue using microscopic approaches to  
Kondo lattice 
models~\cite{Si-Nature,GrempelSi,ZhuGrempelSi,SiZhuGrempel,SunKotliar1,SunKotliar2},
which have provided the basis for local quantum criticality. In this picture,
Kondo screening turns critical at the magnetic quantum critical point.
Here, 
we put the microscopic work in a more general context by considering
the global magnetic phase diagram of the Kondo lattice. We
show that
there exist a number of metallic Fermi liquid phases, which are characterized
by distinct Fermi surfaces. Different classes of QCP naturally
occur
in this phase diagram. Other theoretical approaches to the problem 
of quantum critical
heavy fermions can be found  in Refs.~\cite{Coleman,Pepin,Coleman2,Senthil}.

\section{Global Magnetic Phase Diagram}

We will focus on
the
Kondo lattice model,
\begin{eqnarray}
H &=&  H_f +  H_c + H_K .
\label{kondo-lattice}
\end{eqnarray}
The Hamiltonian for the $f-$electron local moments is
\begin{eqnarray}
 H_f &=&
\frac{1}{2}
\sum_{ ij}
I_{ij}^a
~S_{i}^a
~ S_{j}^a .
\label{H-f}
\end{eqnarray}
Here, $a=x,y,z$ are spin projections, and 
$I_{ij}^a$ is the RKKY interaction between
the spin-$1/2$ moments (one per site).
We use $I$ to label the typical RKKY interaction (say, the dominant
component of the
nearest-neighbor interactions), which is antiferromagnetic.
In addition, $G$ describes the degree of frustration
({\it e.g.} $G=I_{\rm nnn}/I_{\rm nn}$, the ratio of the
antiferromagnetic next-nearest-neighbor interaction over the nearest
neighbor one), or the degree of spatial anisotropy. For our purpose,
it's adequate to know that increasing $G$ corresponds to a decrease
in the strength of the N\'{e}el order. 
\begin{eqnarray}
 H_c &=& 
\sum_{\bf k \sigma} \epsilon_{\bf k}
c_{{\bf k}\sigma}^{\dagger} c_{{\bf k}\sigma} 
\label{H-c}
\end{eqnarray}
describes a band of conduction electrons -- $x$ per site
,
with $0<x<1$ without loss of generality. The bandwidth of the 
conduction electron is $W$.
The two components interact with each other through
\begin{eqnarray}
H_K 
&=& \sum_i J_K ~{\bf S}_{i} \cdot {\bf s}_{c,i} ,
\label{H-K}
\end{eqnarray}
where the Kondo interaction, $J_K$, is antiferromagnetic.

\begin{figure}[!ht]
\begin{center}
\includegraphics[width=0.45\textwidth]{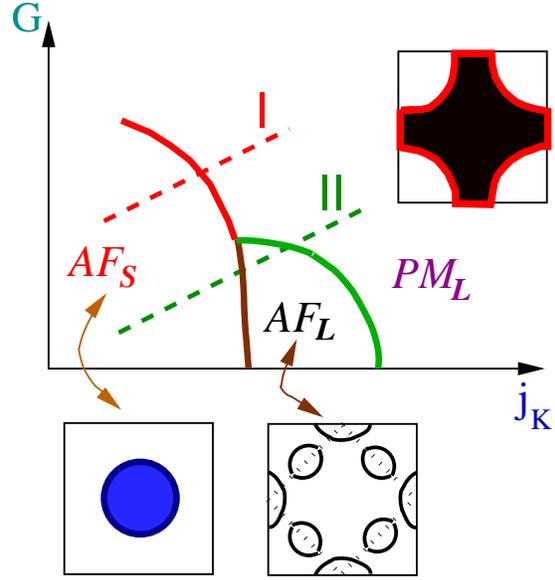}
\end{center}
\caption{The global magnetic phase diagram of the Kondo 
lattice at zero temperature.
$j_K$ is the Kondo coupling measured in terms of
the conduction-electron bandwidth.
$G$ labels
frustration.
As illustrated,
three phases, ${\rm AF_S}$, ${\rm AF_L}$ and ${\rm PM_L}$,
have distinct Fermi surfaces.
The dashed lines
``$I$'' and ``$II$'' label two types 
of
transitions. More detailed descriptions are given
in the main text.}
\label{gpd}
\end{figure}

The
zero-temperature
phase diagram can be specified in the multi-dimensional parameter 
space of $x$, $I/W$, $J_K/W$, and $G$. In a given material, 
the conduction electron density $x$ is fixed, but the other 
parameters can be varied. Here, we will consider a fixed and 
(as in real materials)
relatively 
small $I/W$. 

In Fig. \ref{gpd}, the horizontal axis labels 
$j_K \equiv J_K/W$, 
while the
vertical axis describes the local moment magnetism that is completely
decoupled from the
conduction electrons. When $G$ is sufficiently
large, 
the
conventional N\'{e}el
state becomes unstable
towards
states which
preserve spin-rotational invariance but is translational-invariance 
breaking (spin Peierls) or preserving (spin liquid). We will not get 
into that regime
, but will instead focus
on the region of $G$ where
the local moment component itself remains in the N\'{e}el state.
Still, 
incorporating the parameter $G$
allows us to discuss the phase diagram
beyond the traditional picture~\cite{Doniach,Varma}, which 
arises from considering only an energetic competition between
the RKKY ($I$) and Kondo ($J_K$) couplings.

\begin{figure}[!ht]
\begin{center}
\includegraphics[width=0.45\textwidth]{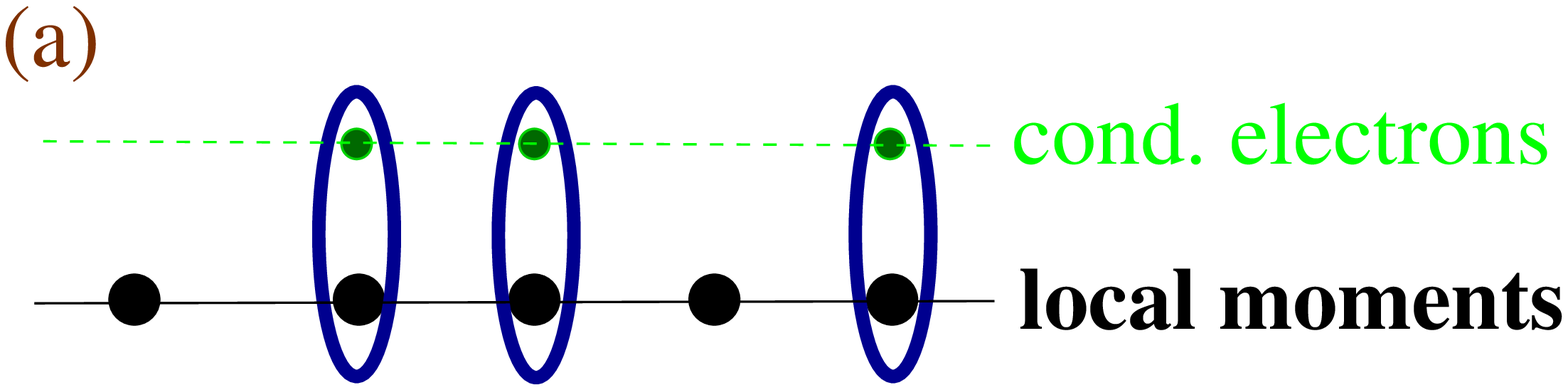}
\vskip 0.8cm
\includegraphics[width=0.45\textwidth]{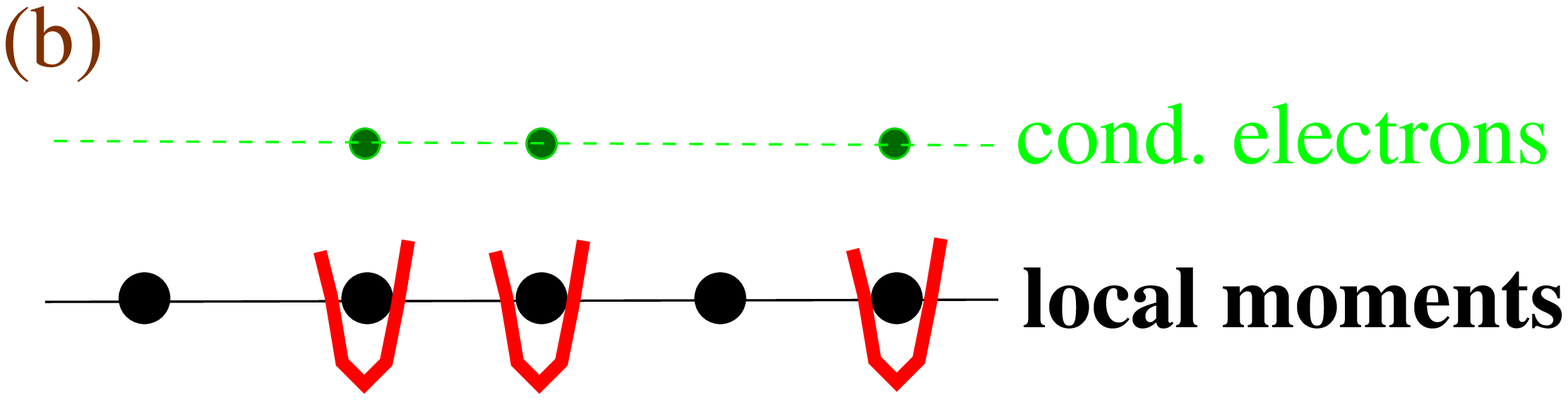}
\end{center}
\caption{(a) Kondo singlets and lone 
local
moments in the 
$J_K \gg W \gg I$ limit;
(b)
In the opposite $J_K \ll I \ll W $ limit,
static Kondo singlets do not form,
but dynamical singlet correlations do exist.
}
\label{grip}
\end{figure}

The magnetic phase diagram is shown in Fig. \ref{gpd}.
The ${\rm PM_L}$ phase describes a heavy Fermi
liquid with a Fermi surface that encloses $1+x$ electrons per
unit cell within the paramagnetic Brillouin zone~\cite{Bickers}.
This phase
can be most easily seen at $J_K/W \gg 1$, as illustrated 
in Fig. \ref{grip}a. 
At each of the $xN_{\rm site}$ sites (where $N_{\rm site}$
is the number of unit cells in the system),
a local moment and a conduction electron form a
tightly bound
singlet,
\begin{eqnarray}
|s>_i = (1/\sqrt{2})(|\uparrow>_f|\downarrow>_c
-
|\downarrow>_f|\uparrow>_c ),
\label{tight-singlet}
\end{eqnarray}
with a large binding energy of order $J_K$.
Each of the remaining $(1-x)N_{\rm site}$ sites hosts a lone
local
moment
which, when projected to the low energy subspace, is written as 
\begin{eqnarray}
|{\rm lone~
local~
moment}~\sigma>_i = (-\sqrt{2}\sigma) c_{i,\bar{\sigma}}|s>_i .
\label{lone-moment}
\end{eqnarray}
In other words, if we consider the $|s>$ as the vacuum state, 
a lone
local
moment behaves as a hole with infinite repulsion (there is only
one conduction electron in the singlet) but with a kinetic energy
of order $W$~\cite{LaCroix}. In the paramagnetic
phase, we can invoke the Luttinger theorem to conclude that the Fermi
surface encloses $(1-x)$ holes or, equivalently,
$(1+x)$ electrons per unit cell. This is the heavy fermion state in which
local moments, through an entanglement with conduction electrons, 
participate in the
electron fluid~\cite{Bickers}.
The Fermi surface is large in this sense,
and the phase is labeled as ${\rm PM_L}$.

Another corner of the phase diagram where exact statements can be made 
is for 
$J_K/W 
(
\ll I/W
)
\ll 1$. For simplicity, we will consider the case
with Ising spin anisotropy. The spin excitation spectrum of the
N\'{e}el ordered local moment component is gapped; it follows that the 
Kondo coupling is irrelevant in the 
renormalization group sense. This is pictorially illustrated in 
Fig.~\ref{grip}b, where the Kondo coupling provides
dynamical
singlet correlation, but does not succeed in forming
any
``grip''
(static Kondo singlet).
Local moments stay charge neutral, and they do not contribute to the
electronic excitations.
The Fermi surface is small. We call this phase ${\rm AF_S}$.

As the system moves from the ${\rm AF_S}$ phase to the 
${\rm PM_L}$ phase, two things happen. First, the singlet 
correlation between the local moments and conduction 
electrons becomes stronger: when the ``grip'' finally forms,
Kondo screening is realized. Second, the 
N\'{e}el order becomes more fragile and eventually goes away.

Traditionally, it is believed that the Kondo screening develops
before the N\'{e}el order disappears. There is then an 
intermediate magnetically ordered state, in which the Fermi volume
in the magnetic zone (for commensurate order) is the same as 
that of the ${\rm AF_S}$ phase. Nonetheless, the Fermi surface 
of this intermediate phase is labeled as large,
in the sense that local moments have become 
a
part of the electron fluid. 
This
${\rm AF_L}$ phase can be thought of as a spin-density-wave
(SDW) state formed out of the heavy fermion quasiparticles
of the ${\rm PM_L}$ phase;
indeed, the Fermi surface of 
the ${\rm AF_L}$ phase is adiabatically connected to that of 
${\rm PM_L}$, when the magnetic order parameter is switched
off. This Fermi surface of the ${\rm AF_L}$ phase 
has a different topology from that
of 
the ${\rm AF_S}$ phase (as can be most easily seen near 
the
multicritical
point);
the two phases are separated by a Lifshitz transition. 
The magnetic quantum transition is between
the ${\rm AF_L}$ and ${\rm PM_L}$ phases, and is labeled type II.

It is also possible, however, for a direct transition
between the ${\rm AF_S}$ and ${\rm PM_L}$ phases.
This is the type I transition shown in Fig.~\ref{gpd}.

\section{Quantum Critical Points}

At the type II magnetic transition,
the effective Kondo screening scale of the lattice -- the coherence
temperature -- is finite. The quantum critical point belongs
to the Hertz-Moriya-Millis type~\cite{Hertz76,Moriya,Millis}.

The type I magnetic transition, however, goes directly from
the ${\rm AF_L}$ phase to the ${\rm PM_L}$ phase.
The transition is second order if 
$z_L$, the quasiparticle residue of the large Fermi surface in
the ${\rm PM_L}$ phase, and $z_S$, its counterpart of the small Fermi
surface in the ${\rm AF_S}$ phase, go to zero as the transition is 
reached from
respective sides.
The coherence temperature vanishes 
-- and 
the
Kondo singlets
disintegrate -- as the QCP is approached from
the ${\rm PM_L}$ side.

At such a magnetic QCP, the destruction of Kondo screening coincides with 
the onset of magnetic ordering. The understanding of
actually
how the quasiparticles
are destroyed at the QCP comes from microscopic considerations.
One mechanism 
is
the local quantum 
criticality~\cite{Si-Nature,GrempelSi,ZhuGrempelSi,SiZhuGrempel}. 

Fluctuations 
of the magnetic order parameter are the softest at the magnetic QCP.
These slow fluctuations in turn decohere the Kondo screening, making
the Kondo effect critical. The latter characterizes the 
emergent
non-Fermi
liquid critical excitations, which are in addition to the critical
fluctuations of the magnetic order parameter.

The local QCP has a number of characteristics.
Electronically,
the $f-$electrons turn from being itinerant to being localized across
the QCP. There are two corollaries. The Fermi surface undergoes a sudden
reconstruction at the QCP. In addition, the continuous vanishing of
both
$z_L$ and $z_S$ implies that the effective mass
diverges as the QCP is approached from both the paramagnetic and 
magnetic sides. It is worth expanding on this feature for the 
magnetic side. The mass enhancement in heavy fermions has traditionally
been associated with the formation of Kondo resonance. How can the
${\rm AF_S}$ phase, having no Kondo resonance, acquire small quasiparticle
residue and large effective mass? As Fig.~\ref{grip}b illustrates,
here, even though Kondo singlet is not formed in the static sense, 
dynamical
singlet correlation 
does occur
and becomes 
stronger as the QCP is approached.
It is this dynamical effect that enhances
both 
the thermodynamic
mass (as measured in
,{\it e.g.},
specific heat coefficient)
and 
electronic
mass (as measured
in, {\it e.g.},
dHvA).

A second feature of the local QCP arises in the magnetic dynamics.
In contrast to the Gaussian fixed point of the
$T=0$
SDW transition,
where $\omega/T$ scaling is violated~\cite{Hertz76,Moriya,Millis},
the interacting nature of the 
local QCP
produces
an $\omega/T$ scaling. Moreover, the magnetic
dynamics contains a fractional exponent. The dynamical spin 
susceptibility turns out to have the 
form~\cite{Si-Nature,GrempelSi,SiZhuGrempel}
\begin{eqnarray}
\chi({\bf q},\omega ) = 
\frac{\rm const.} 
{I_{\bf q} - I_{\bf Q} + (-i\omega)^{\alpha}
M
(\omega/T)} ,
\label{chi-q-omega}
\end{eqnarray}
where ${\bf Q}$ is the antiferromagnetic ordering wavevector. 

\section{Experiments}

Experimental data
in heavy fermions
suggest that both the antiferromagnetic and 
paramagnetic phases are
indeed
Fermi liquids. In YbRh${\rm _2}$Si${\rm _2}$, for 
instance, the resistivity is $T^2$ on both sides of the QCP~\cite{Custers}.
Related features have been observed
in CePd${\rm _{2}}$Si$_{\rm _2}$~\cite{Grosche,Flouquet}
and
CeCu${\rm _{6-x}}$Au${\rm _x}$~\cite{Lohneysen}.
In the
case of CeCu${\rm _{6-x}}$Au${\rm _x}$,
for $x>\sim x_c$ with small $T_N$,
however, the specific heat coefficient
does not appear to saturate at the lowest measured temperatures;
this region remains to be clarified.

There are also extensive Fermi surface measurements via dHvA.
It is well established that the paramagnetic metal phase has
a large Fermi surface~\cite{Lonzarich}. 
Perhaps less well known is the fact that antiferromagnetic heavy fermions
are typically found to have a small Fermi surface
(for
recent 
reviews, 
see Ref.~\cite{Julian}).
Since a large magnetic field
-- which is a big perturbation to heavy fermions --
is necessarily involved
in the experiment
,
it is natural that the 
dHvA
measurement
generically
probes
the parts of the phase diagram 
sufficiently away from the magnetic-transition region. By extension,
it is natural that the ${\rm AF_S}$ and ${\rm PM_L}$ phases are 
the ones that are commonly identified in such measurements.

We now turn to experiments which zoom in on the transition region.
Consider first the inelastic neutron scattering experiments.
CeCu${\rm _{6-x}}$Au${\rm _x}$, at $x =0.1 \approx x_c$, is the 
most striking case of a single crystal showing a dynamical spin
susceptibility with a fractional exponent and an $\omega/T$ scaling,
of the form given in Eq.~(\ref{chi-q-omega})~\cite{Schroder,Stockert}.

Recent measurements~\cite{Kadowaki} have been carried out in 
Ce(Ru${\rm _{1-x}}$Rh$_{\rm x}$)$_{\rm 2}$Si${\rm _2}$. This 
single crystal displays a paramagnetic to antiferromagnetic
metal transition at $x_c \approx 0.04$. Close to this concentration,
the  inelastic neutron
scattering data
is
well described 
by the Lorentzian form,
$\chi({\bf q},\omega ) = {\rm \chi_{\bf q}(T)}/ 
{[1-i\omega / \Gamma_{\bf q}(T)]}$,
with $\Gamma_{\bf Q} \sim T^{3/2}$.
This form, violating $\omega/T$ scaling,
is what is expected in a 3D AF SDW QCP~\cite{Hertz76,Moriya,Millis}.
In addition,
the electrical resistivity and specific heat data are 
also reasonably consistent with the SDW picture.
While it will be important for future experiments to map out the $T_N$
line closer to the $T=0$ transition (the lowest finite $T_N$ that has
been determined so far is of the order of 3 K), the evidence seems
quite strong that we are finally seeing a Hertz QCP!
Unlike the
quasi-2D nature~\cite{Stockert,Schroder} seen in 
CeCu${\rm _{6-x}}$Au${\rm _x}$, the magnetic fluctuations in
Ce(Ru${\rm _{1-x}}$Rh$_{\rm x}$)$_{\rm 2}$Si${\rm _2}$
are
three-dimensional~\cite{Kadowaki}. This
makes
Ce(Ru${\rm _{1-x}}$Rh$_{\rm x}$)$_{\rm 2}$Si${\rm _2}$
to be located
at the lower part of our global
phase
diagram (Fig.~\ref{gpd})
than
CeCu${\rm _{6-x}}$Au${\rm _x}$.
This placement
is consistent with
the identification of
type II and type I
quantum
transitions in these
two materials, respectively.

Consider next
electronic measurements in the immediate
vicinity of the transition. Detailed Hall effect 
studies~\cite{Paschen}
have 
been carried out 
in YbRh${\rm _2}$Si${\rm _2}$.
In this
material,
the anomalous
Hall component is relatively
small
at low temperatures,
allowing the 
extraction of the normal Hall component. The Hall coefficient
shows a rapid crossover at finite temperatures, extrapolating
to a jump in the $T=0$ limit at the magnetic QCP.
The result
provides
strong evidence that
the second order quantum transition in
YbRh${\rm _2}$Si${\rm _2}$
goes
directly from ${\rm AF_S}$ to ${\rm PM_L}$.

We 
already
mentioned that the large magnetic field needed in dHvA 
makes it 
generally
difficult to use this method
to
zero in on the quantum 
critical
point.
A fortuitous situation arises in CeRhIn$_{\rm 5}$.
A magnetic field, of the order used in the dHvA measurement, 
is just what is needed to entirely suppress superconductivity
and expose a pressure induced zero-temperature transition from 
an antiferromagnetic metal to a paramagnetic metal~\cite{Park}. 
Indeed, the dHvA result~\cite{Onuki}
can be interpreted in terms of a sudden reconstruction
of
Fermi
surface,
from
that
of ${\rm AF_S}$ to
its counterpart 
of ${\rm PM_L}$, across
the
critical pressure.
Moreover, the (electronic) dHvA mass shows a large (more than
10-fold) increase as the QCP is approached. Taken together,
these measurements provide strong evidence for a field-
and pressure-induced type I magnetic QCP in CeRhIn$_{\rm 5}$. 

Finally, some thermodynamic ratios also turn out to be illuminating
in this context.
We have
shown in Ref.~\cite{Zhu03} that the Gr\"{u}neisen
ratio $\Gamma$ -- the ratio of the thermal expansion, $\alpha 
\equiv {1 \over V}
{\partial V \over \partial T}$, over the specific heat, $c_p$ --
has to diverge at any QCP where the control parameter is linearly
coupled to pressure. Scaling implies that, at the QCP, 
$\Gamma \sim 1/T^x$, with the exponent $x=1/z\nu$ (where $z$ is
the dynamic exponent and $\nu$ the correlation length exponent).
Measurement~\cite{Kuchler03} in YbRh${\rm _2}$Si${\rm _2}$ does
indeed find such a divergence. Moreover, the exponent $x \approx 0.7$
is different from the value ($1$) expected from an AF SDW QCP, but
is consistent with the value calculated from the local QCP picture.

\section{Summary and Outlook}

We have shown that two types of magnetic metal phases - ${\rm AF_S}$ and
${\rm AF_L}$ - can occur in Kondo lattices, along with the standard
heavy fermion paramagnetic metal phase ${\rm PM_L}$. This opens up
a new type of magnetic quantum phase transition,
which goes 
directly from
${\rm AF_S}$ to ${\rm PM_L}$. The transition is second order when
quasiparticle residues vanish. At this magnetic QCP, the critical
excitations include not only the fluctuations of the order parameter
but also those associated with a critical Kondo screening. Local quantum
criticality is one form of such type of QCP.

We close with a few general remarks.
The 
global phase diagram
makes it
desirable to systematically
study magnetic quantum transitions in heavy 
fermion metals 
with different
degrees of frustration. For instance, ${\rm YbAgGe}$ has a hexagonal
lattice and its spin interactions may very well be frustrated.
Indeed, the magnetic phase transitions in this material
are rather unusual~\cite{Canfield}.
The venerable ${\rm UPt_3}$ also has a hexagonal lattice and 
it could be instructive to
study
quantum phase
transitions in this material or its relatives.

The prominent role played by the destruction of Kondo screening
in our global phase diagram has other implications.
We may, for instance, replace the N\'{e}el order parameter discussed
so far by a spin glass one. 
We are then led to two
types of quantum spin glass
transitions
.
The type II transition 
(${\rm SG_L}$ to ${\rm PM_L}$) is expected to be described by
a Gaussian fixed point~\cite{Sachdev95,Sengupta95},
with
a violation of $\omega/T$ scaling in the magnetic dynamics.
A type I transition (${\rm SG_S}$ to ${\rm PM_L}$), on the other
hand, can correspond to an interacting fixed point,
yielding an
$\omega/T$ scaling. Recent inelastic neutron scattering 
study~\cite{Dai95}
 near a spin-glass QCP of 
${\rm Sc_{1-x}U_{x}Pd_3}$~\cite{Maple00} does indeed find 
an $\omega/T$ scaling and a fractional exponent, suggesting 
a destruction of Kondo screening at this spin-glass QCP.
The striking similarity of these data with those of 
${\rm UCu_{5-x}Pd_x}$~\cite{Aronson95,MacLaughlin} naturally suggests
that the later too originate from a destruction
of Kondo screening at a spin-glass QCP.

Finally, it is possible that the physics of magnetic quantum
criticality with critical Kondo screening in heavy fermion metals
connects to that of certain quantum critical spin liquid
states
in quantum 
insulating 
magnets~\cite{Senthil03,Si04,Senthil05}. Itinerant systems
such as heavy fermions are inherently spin-$1/2$ systems. This is
in contrast to
insulating
magnetic materials,
in which 
the size of spin is
typically larger than $1/2$,
making
quantum effects less pronounced.
So, perhaps, heavy fermion metals can also play an important role 
in the on-going search for
both
critical
and
stable spin liquid states.

I am particularly gratefully to D. Grempel, K. Ingersent,
S. Kirchner, E. Pivovarov,
S. Rabello, J. L. Smith,
J.-X. Zhu, and L. Zhu
for collaborations in this area, and many colleagues 
for discussions. The work has been partially supported 
by NSF Grant No.\ DMR-0424125 and the Robert A. Welch
Foundation.

\end{document}